\documentstyle[psfig,draft,aps]{revtex}

\def\ftrho{F^{(3)}_\rho}
\def\ferho{F^{(8)}_\rho}
\def\ftomega{F^{(3)}_\omega}
\def\feomega{F^{(8)}_\omega}
\def\ftphi{F^{(3)}_\phi}
\def\fephi{F^{(8)}_\phi}
\begin{document}
\draft
\title{Isospin-Breaking Vector Meson Decay Constants From \\
Continuous Families of Finite Energy Sum Rules}
\author{Kim Maltman\thanks{e-mail: maltman@fewbody.phys.yorku.ca}}
\address{Department of Mathematics and Statistics, York University, \\
          4700 Keele St., Toronto, Ontario, CANADA M3J 1P3 \\ and}
\address{Special Research Center for the Subatomic Structure of Matter, \\
          University of Adelaide, Australia 5005}
\author{C.E. Wolfe\thanks{e-mail: wolfe@fewbody.phys.yorku.ca}}
\address{Department of Physics and Astronomy, York University. \\
          4700 Keele St., Toronto, Ontario, CANADA M3J 1P3}
\maketitle
\begin{abstract}
The isospin-breaking vector meson decay constants are determined
from a QCD sum rule analysis of the
vector current correlator $\langle O\vert T\left(
V^3_\mu V^8_\nu \right)\vert O\rangle$, using
a recently proposed implementation of the finite energy sum rule
approach.  The analysis employs the 
three-loop version of the OPE, and two different families of
weight functions.  It is shown that the
requirement of consistency between results obtained 
using these two different
weight families leads to a rather good determination of the parameter
describing the deviation of the $D=6$ condensate term in the OPE
from its vacuum saturation value, and that 
the ability to determine this value has
non-trivial numerical consequences to the analysis.
The phenomenological relevance of
the results to the extraction of the strange quark
mass and the determination of the $6^{th}$
order chiral low energy constant, $Q$, is also briefly discussed.
\end{abstract}
\pacs{}

\section{Introduction}
Because the neutral ($a=3,8$) members of the 
$SU(3)_F$ octet of vector currents,
$J^a_\mu = \bar{q}\gamma_\mu {\frac{\lambda^a}{2}}q$
(with $\lambda^a$ the usual Gell-Mann matrices),
couple to fermions in the Standard Model, it is possible
to use experimental data on the spectral functions
associated with correlators involving these currents
to determine certain quantities of phenomenological
interest.  For example, defining
the scalar correlators, $\Pi^{ab}(q^2)$, by means of
\begin{equation}
i\int\, d^4x\, \exp (iqx)\, \langle 0\vert T\left(
J^a_\mu (x) J^b_\nu (0)\right) \vert 0\rangle \equiv
\left( q_\mu q_\nu -q^2 g_{\mu\nu}\right) \Pi^{ab}(q^2 ),
\label{scalarcorrelator}
\end{equation}
and the corresponding spectral functions, $\rho^{ab}(q^2)$, as usual,
by $\rho^{ab}(q^2)={\frac{1}{\pi}}{\rm Im}\, \Pi^{ab}(q^2)$,
one finds that (1) integrating the difference 
$\rho^{33}(q^2)-\rho^{88}(q^2)$ 
with the weight
function occuring naturally (due to kinematics) in the finite energy sum rule
(FESR) treatment of hadronic $\tau$ decays\cite{taurefs}
produces a sum rule from which one can, in principle,
determine the running strange quark mass, $m_s(\mu )$\cite{narisonms}, and
(2) integrating the same difference $\rho^{33}(q^2)-\rho^{88}(q^2)$
with weight function $w(s)=1/s$ produces a sum rule from which one
can extract one 
of the $6^{th}$ order low-energy constants (LEC's), $Q$, appearing 
in the $6^{th}$ order version of the effective chiral Lagrangian\cite{gk96}.
(See Ref.~\cite{gl85} for a discussion of chiral perturbation theory
(ChPT) and the method of effective chiral Lagrangians in general,
Ref.~\cite{fsq6} for the form of the ${\cal O}(q^6)$ terms in
the effective Lagrangian
in the most general case,
and Ref.~\cite{gk95} for both a discussion of the subset of these terms
surviving when one restricts one's attention
to vacuum correlators, and a definition of $Q$.)

Of course, $J^3_\mu$ and $J^8_\mu$ do not couple separately in the
Standard Model, but only in the combination 
\begin{equation}
J^{EM}_\mu =J^3_\mu +{\frac{1}{\sqrt{3}}}J^8_\mu 
\label{EMcurrent}
\end{equation} 
which gives the light quark ($u,d,s$) part of the electromagnetic (EM) current.
Thus, what is measured in $e^+e^-\rightarrow hadrons$, is 
not the desired quantities,
$\rho^{33}$ and $\rho^{88}$, separately, but the combination
\begin{equation}
\rho^{EM}(q^2)=\rho^{33}(q^2)+{\frac{2}{\sqrt{3}}}\rho^{38}(q^2)
+{\frac{1}{3}}\rho^{88}(q^2)\ .
\label{EMrho}
\end{equation}
In the isospin symmetry limit, $\rho^{38}$ would vanish and, since
one could then classify the final hadronic states according to their
G-parity, it would be straightforward to separate
the isovector ($33$) and isoscalar ($88$) components of
the EM spectral function.  

In the presence of isospin breaking, however,
this process is no longer so straightforward.  The most obvious
experimental signature of the presence of isospin-breaking in
$e^+e^-\rightarrow hadrons$ is the interference shoulder in the
$e^+e^-\rightarrow \pi^+\pi^-$ cross-section in the $\rho$-$\omega$
region\cite{barkovetc}.  
The $e^+e^-\rightarrow \omega\rightarrow \pi^+\pi^-$ contribution
to $\rho^{EM}$ is clearly, to leading order in isospin-breaking,
to be associated with $\rho^{38}$ and hence is usually removed
explicitly in analyzing the data.  This removal is accomplished by
(1) fitting the parameters of
a model for the total $e^+e^-\rightarrow \pi^+\pi^-$
amplitude, consisting of $\rho$, $\omega$ and possible
background contributions, to the experimental data, (2) removing the
$\omega$ contribution once the fit has been performed and (3) squaring
the modulus of the
remaining $\rho$ contribution and identifying
this result with the $\rho$ contribution to $\rho^{33}$\cite{hocker}.  

While this procedure
{\it does} remove one source of isospin breaking contamination
from the nominal
$\rho^{33}$ so extracted, it is easy to see that other such contaminations
still remain.  Indeed, once one allows
isospin-breaking, the physical $\rho$ and $\omega$ are admixtures of pure
isovector and isoscalar states, the size of the admixture of
the ``wrong'' isospin 
component being governed by the scale of isospin breaking.
As a consequence,
the intermediate $\rho$ contribution to $\rho^{38}$, for example,
does not vanish.  In fact, if one denotes the pure isovector
$\rho$ state by $\rho^{(0)}$ and the pure isoscalar $\omega$ state
by $\omega^{(0)}$, one expects $\rho$ contributions to $\rho^{38}$
from two sources: (1) that due to $\rho^{(0)}$-$\omega^{(0)}$ mixing
(a one particle reducible contribution, with coupling of the
isovector current to the $\rho^{(0)}$ component and the isoscalar
current to the $\omega^{(0)}$ component of the $\rho$),
and (2) that due to the ``direct'' (one particle
irreducible, $1PI$) coupling of the $\rho^{(0)}$ component to the
isoscalar current (such a coupling being unavoidable in any hadronic
effective Lagrangian based on QCD).  
Thus, removing the contribution due to the intermediate
state $\omega$ from the $e^+e^-\rightarrow \pi^+\pi^-$ cross-section, 
while removing part of the $\rho^{38}$ contribution, does not remove
it all.  One is then left with,
{\it not} the desired quantity, $\rho^{33}$, but rather with a combination of
$\rho^{33}$ and the residual part of $\rho^{38}$ associated with
the intermediate $\rho$ state (plus possible additional
such contaminations from elsewhere in the spectrum).  
Similar isospin-breaking flavor $38$
contributions exist for $e^+e^-\rightarrow \omega\rightarrow 3\pi$,
complicating the extraction of the isoscalar spectral function.

Corrections for such isospin breaking effects, {\it which are unavoidable so
long as no process exists in which only one of the two neutral
flavor currents couples}, are thus necessary if one
wishes to perform phenomenological analyses of the type mentioned
above.  Such corrections would also be important in performing
precision tests of CVC, which involve a comparison of
$\rho^{33}$ and the charged
isovector spectral function $\rho^{(+)}$, measured in hadronic
$\tau$ decays (see, for example, Ref.~\cite{ALEPH}).  

It is easy to see that, to be able to make these corrections (at least
in the region below $s\sim 2$ GeV$^2$, where the EM spectral function
is, experimentally, resonance dominated), it is sufficient to determine
the isospin-breaking vector meson decay constants.  
Let us first
clarify notation.  We define the flavor $3$ and $8$ vector meson
decay constants via
\begin{equation}
\langle 0\vert J^a_\mu \vert V(k)\rangle = m_V F^a_V\, \epsilon_\mu (k)
\end{equation}
where $V=\rho ,\omega ,\phi ,\cdots$,
$\epsilon_\mu (k)$ is the vector meson polarization vector,
and $a=3,8$. $\ftrho$, $\feomega$ and $\fephi$ are non-zero in the limit of
isospin symmetry; $\ferho$, $\ftomega$ and $\ftphi$ zero in
the absence of isospin breaking.  The experimentally determined
EM decay constants, $F^{EM}_V$, are then given by
\begin{equation}
F^{EM}_V=F^3_V+{\frac{1}{\sqrt{3}}}F^8_V\ .
\label{femphysical}
\end{equation}
Thus, for example, the broad $\rho$ contribution to $\rho^{EM}$,
usually taken to be associated purely with $\rho^{33}$, consists not only of
a flavor $33$ contribution proportional to $\left[ F^3_\rho\right]^2$,
but also of a flavor $38$ contribution proportional to 
${\frac{1}{\sqrt{3}}}F^3_\rho F^8_\rho$.  The $\omega$
contribution to $\rho^{EM}$, similarly, 
contains both a flavor $88$ part proportional
to $\left[ F^8_\omega\right]^2$ and a flavor $38$ part proportional
to ${\frac{1}{\sqrt{3}}}F^3_\omega F^8_\omega$.  The flavor $38$
parts, in both cases, are present only due to isospin breaking, and
have to be removed from the experimental $\rho$ and $\omega$ 
contributions to $\rho^{EM}$ in order to obtain the corresponding
$\rho$ contribution to $\rho^{33}$ and $\omega$ contribution to
$\rho^{88}$.

It is important to stress
at this point that the conventional ``few-percent'' rule-of-thumb
for estimating the size of isospin-breaking effects, which
might lead one to expect such effects to be numerically negligible,
is inapplicable
in the cases involving $\rho^{33}(q^2)-\rho^{88}(q^2)$
discussed above.
This
is true for a number of reasons.  First, because the 
difference of spectral functions is itself 
flavor-breaking, 
the {\it relative} importance of isospin breaking is enhanced by
a factor $\sim 3$, characteristic of the inverse of the 
scale of flavor-breaking.
Second, the
effect of $\rho$-$\omega$ mixing naturally produces corrections
for the $\rho$ contribution to $\rho^{33}$ and 
$\omega$ contribution to $\rho^{88}$ which are opposite in sign; 
the effects therefore add
when the difference is taken.  Finally,
there is a natural numerical enhancement which makes the 
size of the correction needed to remove
the $\rho^{38}$ part of the $\omega$ contribution to $\rho^{EM}$,
and hence isolate $\rho^{88}$, larger than naively expected\cite{krmspec}.
The latter two points are discussed in somewhat more detail in Section II
below.

In what follows, we evaluate the isospin-breaking vector meson
decay constants by performing a QCD sum rule analysis of the 
isospin-breaking vector current correlator $\Pi^{38}$.  The vector
meson spectral contributions are, in this case, proportional to
$F^3_V F^8_V$, so that a determination of this product, in combination
with the experimental determination of $F^{EM}_V$, given in terms
of $F^3_V$ and $F^8_V$ above, allows a separate determination of
$F^3_V$ and $F^8_V$.  The rest of the paper is organized as follows.
In Section II we discuss qualitative expectations for the pattern
of isospin-breaking corrections
based on the
structure of the leading (chiral) order terms in the vector meson
effective chiral Lagrangian, as well as semi-quantitative
expectations for their probable
scale which can, using this perspective, be obtained from
experimental data.  In Section III, we discuss
briefly the form of QCD sum rules employed (a version of FESR), and
the advantages of this approach.  In Section IV, we discuss the
input used for the hadronic and OPE sides of the sum rules employed, and
present our results.
Some advantages of the approach, in particular {\it viz a vis} the handling
of the $D=6$ terms in the OPE of the $38$ correlator, will also
be discussed here.  Finally, in Section V we summarize, and make some
brief comments on the phenomenological significance of our results.

\section{Chiral Constraints and the Scale of Isospin-Breaking Corrections}
ChPT provides both an underlying conceptual
framework and systematic procedure\cite{ccwz}
for writing down the most general effective Lagrangian relevant to 
a given set of hadronic states which fully incorporates the symmetries
of QCD and implements the broken symmetries (such as chiral symmetry)
with the same pattern of symmetry breaking as occurs in QCD.  
Although the resulting effective Lagrangian, ${\cal L}_{eff}$, 
is non-renormalizable, and hence contains a infinite number of
low-energy constants (LEC's) (coefficients of those terms allowed
by the symmetry arguments that go into constructing ${\cal L}_{eff}$),
it is possible to formulate the theory
in such a way that 
only a finite number of such terms appear to a given order in the
chiral, or low-energy expansion.  (For so-called ``heavy'' fields,
those whose masses are non-vanishing in the chiral limit, this
requires a re-formulation in terms of velocity-dependent 
fields\cite{heavybaryon,heavyvector}.)
The leading order terms in this
expansion (in which light quark masses, $m_q$, $q=u,d,s$, count
as ${\cal O}(q^2)$, with $q$ representing some soft external momentum),
incorporate the leading constraints associated either with chiral
symmetry, or the symmetry pattern of its breaking.  

The heavy field implementation of ChPT given in Ref.~\cite{heavyvector}
for the vector mesons and their interactions with the members of the
pseudo-Goldstone boson pseudoscalar octet
provides some useful information about the pattern of isospin-breaking
mixing in the vector meson sector.  One must bear in mind that, in
general, mixing in field theory is more complicated than in Hamiltonian
quantum mechanics.  This is because, in an effective field theory,
one in general has (and cannot, in fact, avoid) momentum-dependent
mixing terms.  This means that there are off-diagonal elements of the
wave-function renormalization matrix and, as a consequence, the
relation between the original, unmixed fields and the final, renormalized
and diagonalized fields, involves the product of a symmetric matrix and
a rotation (and hence is not itself a rotation).  (See, for example, the
discussion of the effects of $\pi^0$-$\eta$ mixing in Ref.~\cite{gl85};
an expanded version is also given as part of the discussion of
the treatment of the mixed-isospin axial current correlator in
Ref.~\cite{kma3a8}.)
From the heavy vector meson
field formulation, however, one sees that the leading (in chiral order)
term in ${\cal L}_{eff}$ generating isospin-breaking mixing involves
one power of the quark mass matrix and no derivatives\cite{heavyvector}, 
and hence
produces no off-diagonal contributions to the wave-function renormalization
matrix.  The {\it leading order} mixing effect thus results in a 
physical $\rho$ and $\omega$ basis which is related by a rotation
to the original pure isospin $\rho^{(0)}$, $\omega^{(0)}$ basis, just
as in the quantum mechanical case.  The ``wrong'' isospin
$\omega^{(0)}$ admixture in the physical $\rho$ state is thus equal
in magnitude, but opposite in sign,
to that of the $\rho^{(0)}$ admixture in the physical $\omega$ state,
at this order in the chiral expansion.
While higher (chiral) order corrections exist, this pattern should
remain approximately valid, even at higher order.

Let us now consider the vector meson decay constants.  Since the
chiral limit is also $SU(3)_F$ symmetric, the leading order 
term generating the vector meson decay constants is necessarily
$SU(3)_F$-symmetric.
When one now considers
the effects of flavor- and isospin-symmetry breaking (recalling
that both are generated by the quark mass matrix, and hence both
are produced by the same set of terms in the effective Lagrangian),
there are two potential sources of flavor and isospin breaking in the vector
meson decay constants.  The first is that associated with higher
order terms, involving at least one power of
the quark mass matrix, coupling the external photon field to
the vector meson nonet, the second that induced by the leading
quark-mass-dependent term, responsible for mixing, discussed above.
The leading order mixing effect simply reproduces the
standard leading order $SU(3)_F$ mixing analysis\cite{heavyvector},
leading to near ideal mixing in the vector meson sector.
As is well-known, the combination of ideal mixing and neglect of
flavor-breaking in the EM couplings of the unmixed states leads
to the prediction that the vector meson EM decay constants, measured
experimentally in $V\rightarrow e^+e^-$,\cite{pdg98}
should be in the proportions $F^{(0)}_\rho :F^{(0)}_\omega :F^{(0)}_\phi
= 3:1:-\sqrt{2}$, where the superscript $(0)$ indicates that the
couplings refer to the ideally mixed, but isospin pure
vector meson states.  That this prediction
is borne out by experiment represents empirical evidence that,
despite the potential $SU(3)_F$-breaking photon coupling contributions
being of the same formal order as effects induced by mixing,
the former are numerically suppressed relative to the latter.
Since flavor breaking and isospin breaking are generated by the same
terms in the effective Lagrangian, this implies that isospin
breaking in the vector meson decay constants should also be dominated
by mixing effects.  

If we take this point of view then, up to sub-leading corrections,
we find, for the physical $\rho$ and $\omega$ decay constants, now
including isospin breaking and taking into account the
relation $F^{(0)}_\rho \simeq 3F^{(0)}_\omega$,
\begin{eqnarray}
F^{EM}_\rho&=&F^{(0)}_\rho -\epsilon F^{(0)}_\omega
\simeq F^{(0)}_\rho\left( 1-\frac{\epsilon}{3}\right) \nonumber \\
F^{EM}_\omega&=&F^{(0)}_\omega +\epsilon F^{(0)}_\rho
\simeq F^{(0)}_\omega\left( 1+3\epsilon \right) \, ,
\end{eqnarray}
where $\epsilon$ is the leading order mixing angle, defined via
\begin{equation}
\rho = \rho^{(0)} -\epsilon\, \omega^{(0)} \, ,\qquad
\omega =\omega^{(0)} +\epsilon\, \rho^{(0)}\ .
\end{equation}
We note two relevant features of these results: (1) because of the
dominance by mixing, the corrections required to convert the
pure isovector $F^{(0)}_\rho$ coupling to the experimental
$F^{EM}_\rho$ coupling is opposite in sign to that required
to convert the pure isoscalar $  F^{(0)}_\omega$ coupling
to the $F^{EM}_\omega$ and, (2) because of the pattern of ideal
mixing and the numerical suppression of the isoscalar current
relative to the isovector current in $J^{EM}_\mu$, the magnitude
of the correction is a factor of $9$ larger in the $\omega$
than in the $\rho$ case.  This does not mean that, somehow,
isospin breaking has become huge in the $\omega$ case, but
rather reflects the expected natural suppression of the $\omega$ coupling
relative to that of the $\rho$, which is confirmed by experiment.

In general, but particularly in light of the large numerical enhancement 
just discussed, it
is crucial to find a means of estimating the isospin-breaking 
contributions to the vector meson EM decay constants.  In view of
the discussion above, although there exist additional 
effects beyond those associated with mixing at
leading order, one can get a rough idea of the
size expected for the isospin-breaking
vector meson decay constants by considering experimental information
on $\rho$-$\omega$ mixing, and ignoring all non-mixing effects.  
Although crude, this estimate provides an additional qualitative
constraint for our later sum rule analysis.

In order to obtain the parameter $\epsilon$ describing $\rho$-$\omega$
mixing at leading order, it is sufficient to determine the 
off-diagonal element, $\Pi_{\rho\omega}$, of the vector meson
self-energy matrix.  In the past, values for $\Pi_{\rho\omega}$
in the range $\sim -4000$ MeV$^2$ have been quoted, based on
simplified analyses of $e^+e^-\rightarrow \pi^+\pi^-$ data in the 
interference region which effectively assume that the one-particle
irreducible $\omega^{(0)}\pi^+\pi^-$ vertex is zero, even in the presence
of isospin breaking.  Since effective operators which
generate such a coupling exist in the vector meson effective
Lagrangian, however, this assumption is unphysical (in the sense of being
incompatible with QCD).  Once one
includes contributions to the $\omega\rightarrow\pi\pi$ amplitude
generated both by $\rho$-$\omega$ mixing {\it and} the 1PI vertex
(whose strength we will denote by $g_{\omega\pi\pi}^{(0)}$),
the analysis of the experimental data is somewhat more complicated but,
in principle, allows a separate determination of both $\Pi_{\rho\omega}$
and the isospin-breaking ratio of couplings of the isospin pure states
$G=g_{\omega\pi\pi}^{(0)}/g^{(0)}_{\rho\pi\pi}$\cite{mow,otw}.

An important feature of the analysis framework developed in Ref.~\cite{mow}
is that, from it, one understands that the smallness of previously quoted
errors for $\Pi_{\rho\omega}$ is an artifact of the unphysical
assumption $G=0$, and does not survive the more general treatment.
It is worth outlining the reason why this is the case since, in doing
so, it will become clear that it is a difficult task to improve the
experimental situation sufficiently to really pin down the mixing
contribution.

The contribution of the {\it physical} (i.e., mixed-isospin) $\omega$
to the amplitude for $e^+e^-\rightarrow \pi^+\pi^-$ is obtained
experimentally by determining the timelike pion form factor, $F_\pi (q^2)$,
in the interference region and fitting it to a form
\begin{equation}
F_\pi (q^2)\propto \left[ {\frac{1}{q^2-m_\rho^2}}+{\frac{A^{i\phi}}
{q^2-m_\omega^2}}\right] \, +\, {\rm background}
\label{fpi}
\end{equation}
where $m_V^2$ are the complex pole positions, 
$m_V^2=\hat{m}_V^2-{\rm i}\hat{m}_V\Gamma_V$, and the fit parameter,
$\phi$, is known as the ``Orsay phase''.  The $\omega$ contribution
in Eq.~(\ref{fpi}) is generated by the coupling of the physical
$\omega$ to $\pi^+\pi^-$ which, as discussed above, has two sources:
1PR ($\rho^{(0)}$-$\omega^{(0)}$ mixing), and 1PI (associated with the
$\omega^{(0)}\pi\pi$ vertex).  The physical coupling is given, in
terms of these contributions, by
\begin{equation}
g_{\omega\pi\pi}=g^{(0)}_{\omega\pi\pi}+\epsilon\, g^{(0)}_{\rho\pi\pi}\ ,
\label{gomega}
\end{equation}
where, as usual, the superscript $(0)$ indicates couplings of the unmixed
isospin pure states.  In the (physically plausible) approximation
that one assumes saturation of the imaginary part of $\Pi_{\rho\omega}$
by $\pi\pi$ intermediate states, one finds
\begin{equation}
{\rm Im}\, \Pi_{\rho\omega}({m}_\rho^2)=-G\hat{m}_\rho\Gamma_\rho
\label{impi}
\end{equation}
and hence, in the narrow $\rho$-$\omega$ interference region,
\begin{equation}
\Pi_{\rho\omega}\simeq\tilde{\Pi}_{\rho\omega}-{\rm i}G
\hat{m}_\rho\Gamma_\rho
\label{fullpi}
\end{equation}
where $\tilde{\Pi}_{\rho\omega}$ is now real.
The mixing angle $\epsilon$ is then given by\cite{mow}
\begin{eqnarray}
\epsilon&=&{\frac{\Pi_{\rho\omega}(m_\rho^2)}{m_\omega^2-m_\rho^2}}
\nonumber \\
&=&-{\rm i}z\tilde{T}-zG\ ,
\label{epsilon}
\end{eqnarray}
where
\begin{equation}
z\equiv {\frac{{\rm i}\hat{m}_\rho\Gamma_\rho}{m_\omega^2-m_\rho^2}},
\qquad\qquad \tilde{T}\equiv {\frac{\tilde{\Pi}_{\rho\omega}(m_\rho^2)}
{\hat{m}_\rho\Gamma_\rho}}\ .
\label{zttilde}
\end{equation}
One then finds, upon substitution of Eq.~(\ref{epsilon})
into Eq.~(\ref{gomega}), that
\begin{equation}
g_{\omega\pi\pi}=\left[ G(1-z)+{\frac{\tilde{\Pi}_{\rho\omega}(m_\rho^2)}
{\hat{m}_\rho\Gamma_\rho}}\right] \, g^{(0)}_{\rho\pi\pi}\ .
\label{renard}
\end{equation}

In many places in the literature, one finds the approximation
$m_\omega^2-m_\rho^2\simeq {\rm i}m_\rho\Gamma_\rho$ employed.
The deviation of the quantity $z$ 
from $1$ represents the error made in employing this
approximation.  Since ${\rm Re}\, z\simeq 1$ and ${\rm Im}\, z$
is small ($\sim .2$ to $.3$), one might think it rather safe to set
$z=1$ in the above analysis (this approximation was, in fact,
made uniformly in analyses previous to the discussion of Ref.~\cite{mow}).
If it were true that this approximation were reliable, then the
effect of $G$ in Eq.~(\ref{renard}) would cancel exactly\cite{renardarg},
and the experimental data would determine the real part of
$\Pi_{\rho\omega}$ in the interference region with the usually
quoted errors ($\tilde{\Pi}_{\rho\omega}(m_\rho^2)=-3844\pm 271$ 
MeV$^2$; see Ref.~\cite{otw,twold,cb87} and earlier references cited therein).

Unfortunately, it turns out that the approximation is both
misleading and unreliable.  The reason is that, although
$z$ is approximately real and near $1$, $(1-z)$ is dominantly
imaginary.  Since the denominator of the second term in
Eq.~(\ref{renard}) is also dominantly imaginary, the two terms
add nearly constructively.  Were the phases of these terms to be
actually identical, of course, it would be impossible to use
experimental information to separate them, no matter how
precise that data.  Fortunately,
there is a small phase difference which, at least in principle,
means that a determination, with sufficient accuracy, 
of both the magnitude, $A$, and
phase, $\phi$, of the $\omega$ contribution
to $F_\pi$, 
would allow separate determination of $G$ and $\tilde{\Pi}_{\rho\omega}$.
From this, one would be able to reconstruct $\Pi_{\rho\omega}$ and hence
determine $\epsilon$.  There is, however, a practical difficulty,
which limits the accuracy obtainable using present experimental
data.  This is a result of the fact that the
phase difference of the two terms is not large, so one is
rather close to being in the 
``identical phase'' situation, for which a separate extraction of the
two terms would be impossible.
If one takes the updated numerical
analyses of Ref.~\cite{otw}, for example, one sees that values
of $\tilde{\Pi}_{\rho\omega}$ between $-4000$ and $-8000$ MeV$^2$
are allowed by present data (with a central value $\sim -6800$ MeV$^2$),
and that, while the central extraction for $G$ is moderately
large $\sim .1$, $G=0$ is only $2{\frac{1}{2}}\, \sigma$ distant.
A significant improvement in this situation would require a significant
reduction of the errors in the determination of the Orsay phase.
Unfortunately, the prospects for seeing 
such an improvement in the near future are remote, at present.

Although, with present experimental accuracy, the errors in the
determination of $\epsilon$ are disappointingly large, we 
can, nonetheless, as explained above, use the range of values obtained in
Refs.~\cite{mow,otw} to help set a rough scale for the
expected size of those corrections required to go from $F^{EM}_\rho$
to $F^3_\rho$, and $F^{EM}_\omega$ to $F^8_\omega$.  Using the
central values for the four fits given in Table 1 of Ref.~\cite{otw},
one finds
that $F^3_\rho$ is {\it less than} $F^{EM}_\rho$ by between $0.3$ and $3.8\%$
(the former corresponding to fixing $G=0$ by hand, the latter to
the MOW and A solutions contained in the Table of Ref.~\cite{otw}) and
$F^8_\omega$ {\it greater than} $F^{EM}_\omega$ by between
$2.6$ and $24.6\%$.  We will see that the solutions obtained below
via the sum rule analysis satisfy these rather loose constraints,
and, in fact, provide an alternate determination
which is to be favored since it both has considerably smaller
errors and is free of the
uncertainties associated with effectively truncating at leading
chiral order (which goes into the estimate just discussed).

\section{QCD Sum Rules and the Choice of the FESR Method}
As is well-known, the properties of unitarity and analyticity 
lead to the existence of various (appropriately subtracted) dispersion
relations for typical hadronic correlators, $\Pi (q^2)$.  
By the term ``dispersion relations'' we mean here
both those relations
based on the Cauchy representation theorem, and those based on Cauchy's
theorem (the vanishing of the integral of an analytic function
over a closed contour entirely in the region
of analyticity).  The generic term ``QCD sum rules''
describes those versions of these relations obtained by making
kinematic restrictions which allow one to
take advantage of the asymptotic freedom of QCD.
In the first case, one simply writes
the representation theorem for 
$\Pi (q^2)$ with $q^2$ large and spacelike;
in the second, a choice
of the integration contour is made so
that at least part of the contour
lies in the region of large spacelike momenta.  
In either case, the point of the kinematic restriction is to
take advantage of the fact that, in the region of large
spacelike momenta,
computational techniques based on
the operator product expansion 
(OPE)/perturbative QCD (pQCD) 
are applicable.

The generic form of the sum rules generated using the 
representation theorem is then, up to possible subtractions,
\begin{equation}
\Pi (q^2)=
\int^\infty_{s_{th}}\, ds {\frac{\rho (q^2)}{s-q^2}}\ ,
\label{normsr}
\end{equation}
where the $s_{th}$ is the lowest physical
threshold in the channel in
question and the spectral function, $\rho$, is defined as usual by
$\rho ={\frac{1}{\pi}}{\rm Im}\, \Pi$.
The LHS is to be computed using the OPE/pQCD, while
the RHS is given in terms of measured spectral data and/or some spectral
ans\"atz (involving a limited number of unknown
parameters) for the unmeasured part of the spectral function.
As first pointed out by Shifman, Vainshtein and Zakharov (SVZ)\cite{svz},
the utility of such relations is greatly improved by Borel transformation.
The effect of the Borel transform is (1) to replace the
weight $1/\left( s-q^2\right)$ in the hadronic spectral
integral on the RHS of Eq.~(\ref{normsr}) with
$exp\left( -q^2/M^2\right)$, where $M$, the Borel mass, is a
parameter of the transformation, (2) to kill all subtraction terms
and (3) to create a factorial suppression of the contributions
of higher dimensional operators on the OPE side of the equation
($c/\left( Q^2\right)^n\rightarrow c/(n-1)! M^{2n}$).
Ideally, on the hadronic side, one would like to
choose $M$ as small as possible, in order to suppress the contributions
from the large-$s$ part of the integral, for which
the spectral function will typically be complicated, and difficult
to model.  In contrast, to improve the convergence of the OPE
when truncated at operators of relatively low dimension, one
wants $M$ as large as possible.  To be usable, SVZ sum rules thus require
a compromise:
one must hope to find a ``stability window'' in $M$, i.e., a region of 
$M$ values
for which both contributions from the complicated part of the
spectral distribution (on the hadronic side) and from 
the highest dimension operators (on the OPE side) are not too large.
Typically, because of the compromise nature of the choice of
stability window, this means that neither the contributions
from the large-$s$ part of the spectrum, nor that from the highest
dimension operator retained on the OPE side, are 
negligible\cite{svz,review,book,skeptics}.

For the case of sum rules based on Cauchy's theorem, a common choice
of integration contour
(and the one we will employ in the analysis below) is that
shown in Fig.~1.  The radius, $s_0$, of the circular
part of the contour is to be taken large enough that the OPE,
to the order available, is reliable in the spacelike region of the circle.
The resulting
sum rule is then generically of the form
\begin{equation}
{\frac{-1}{2\pi {\rm i}}}\, 
\oint_C\, dq^2\, w(q^2) \Pi (q^2)\ =\ \int_{s_{th}}^{s_0}\,
dq^2\, w(q^2) \rho (q^2)\ ,
\label{fesrbasic}
\end{equation}
where $w(q^2)$ is any function of $q^2$ analytic in the region of
integration, and
$C$ denotes the circular part of the contour, traversed counterclockwise
(from above to below the cut).  The OPE is then to be
used on $C$ (see also the discussion
to follow), while spectral data and/or a spectral ans\"atz is
to be employed in the ``hadronic'' integral on the RHS.  Such sum rules
are called, generically, finite energy sum rules (FESR's), and
have usually been employed with integer-power weights
($w(s)=s^k$, with $k=0,1,2,\cdots$)\cite{fesrrefs,bpr}, though
the standard theoretical treatment of hadronic $\tau$ decays,
which is of the FESR type, involves a more complicated weight,
as determined by kinematics\cite{taurefs}. 

It is useful, for the discussion which follows,
to recall at this point the distinction
between ``local'' and ``semi-local''
duality. 
For a typical hadronic correlator, $\Pi (q^2)$, as above,
the OPE is expected to be reliable,
not only for $q^2$ large and spacelike, but also for those
$q^2$ on any circle of sufficiently large
radius in the complex $q^2$-plane, apart possibly from 
some
region whose size is hadronic in scale about the timelike real
axis (where the
effects of confinement are expected to be important)\cite{pqw}.  
The term ``local duality'', in this context, refers to the
postulate (underlying, for example,
all statements giving the number of subtractions
required to make a given dispersion relation 
convergent in QCD) that, once one is in a region $q^2\sim s_0$ for
which
the separation between adjacent
resonances is small compared to the typical resonance width,
the region of validity of the OPE on the
circle of radius $s_0$ extends all the way down to the
real timelike axis.  This is equivalent to saying that
the hadronic spectral function can then be calculated
using the OPE.  This is what allows, for example, 
the hadroproduction ratio, $R$,
in $e^+e^-$ scattering to be computed
theoretically in the asymptotic region using perturbative methods.  
Similarly, ``semi-local duality'' refers
to the idea that, at somewhat lower (``intermediate'') scales,
where local duality is no longer valid,
nonetheless, averaged over some range of (timelike)
momenta, the mean values given by using either the actual
hadronic spectral function or
the OPE version thereof should be the same.  It is important
to understand that, empirically, the condition that
resonance spacing be {\it much smaller} than typical resonance widths
is crucial to the validity of local duality.  Indeed, one can
test local duality through the application of various FESR's
in the case of the isovector vector channel, for which
the hadronic spectral function is very accurately measured in
hadronic $\tau$ decays\cite{ALEPH}.  One finds that, even at
scales as large as $m_\tau^2\simeq 3.2$ GeV$^2$, and even though
the experimental spectral function appears rather featureless
in this region, nonetheless, local duality is rather poorly
satisfied\cite{kmfesr}.  (Note that in this
channel, at these scales, the separation
of resonances is comparable to their widths.)

In light of the above discussion, we distinguish
three (timelike) kinematic regimes: ``low'' (the region
of narrow, well-isolated resonances), ``intermediate'' (the
region of the validity of semi-local duality) and ``high''
(the region of the validity of local duality, for which
the hadronic spectral function can be reliably obtained
using pQCD/OPE methods).  The distinction is important because
of the ways, described above, in which sum rules are practically implemented.
Typically, one attempts to use known information, contained in the
OPE (i.e., $a(Q^2)=\alpha_s(Q^2)/\pi$, 
and the values of various
vacuum condensates), to place constraints on the parameters
of the (hopefully, physically-motivated)
ans\"atz for the corresponding unknown (or not fully-known) 
spectral function.  This attempt will, of course, be successful only
if (1) one is able to
work at scales for which the OPE, truncated at operators
of relatively low dimension, and at a given perturbative order for
the Wilson coefficients of these operators, is well-converged,
and (2) one has a {\it qualitative} form for the hadronic
spectral ans\"atz which is both physically plausible and does
not involve a large number of unknown parameters.  The latter
condition can, realistically, be satisfied only if the
scale up to which one needs the spectral ans\"atz is not too far into
the intermediate region.  Once the higher resonances start to
overlap, and various multiparticle background processes start
to become important, the general form of the spectral function
will become increasingly difficult to anticipate in advance
(until, that is, one is at sufficiently large scales that
local duality finally becomes valid).  

In the SVZ version of QCD sum rules,
as explained above, it is only rarely possible to choose 
the stability window for the analysis in
such a way as to avoid contributions
from the region where either one does not know the
qualitative form of the spectral function, or where, even if one
does, to implement it fully would involve 
the use of more free parameters than could
be determined reliably, given the limited amount of information
available in the truncated OPE\cite{skeptics}.
Conventionally, this problem is dealt with 
by employing a spectral ans\"atz in which (1) the low-$s$
region is assumed to be dominated by 
one or two low-lying resonance contributions
and (2) the intermediate- and high-$s$ region
is approximated using the local duality version of the spectral function,
which one assumes to start at some ``continuum threshold'',
$s_0$.  It is
well-known that this form of ``continuum ans\"atz'' 
represents a rather crude approximation,
and hence can create significant uncertainties in the analysis
if the contributions from the continuum part of the spectral
integral are large for $M$ values in the stability window. Typically,
one attempts to minimize this problem by avoiding $M$ values
for which this is the case, but the cost
of doing so is poorer convergence on the OPE side of the
sum rule and hence increased uncertainties associated
with neglect of higher dimension contributions.

A similar problem exists for the integer-power weighted version
of FESR's.  One advantage of the FESR approach is that,
in contrast to the SVZ method, the choice of scale appearing
on the OPE side of the sum rule ($s_0$)
is not constrained by the requirement of working in a stability window.
One is, of course, still forced to choose $s_0$ not
too far into the intermediate
region, since otherwise the spectral ans\"atz would be
too complicated to allow a reliable analysis.  
The possibility of working at such intermediate scales
in the FESR approach can,
nonetheless, 
represent a practical advantage
over the SVZ approach in certain cases.  This is true of those channels
for which the stability window of the SVZ analysis lies at
relatively low $M$ ({\it e.g.}, $M\sim 1$ GeV$^2$, as found
for many applications
in the literature).  In such cases, the
larger scale of the FESR analysis leads to an improvement of both the 
convergence by operator dimension and
convergence by perturbative order of the Wilson coefficients
on the OPE side of the sum rule.
Unfortunately, this advantage is usually more than offset
by an increase in the difficulties associated with the use of the
local duality approximation in the intermediate region.  The
reason is obvious: 
with integer-power weights, the region near the timelike real axis
where the circular part of the contour joins the cut, does not
have the exponential suppression present for the ``continuum''
contributions in the SVZ approach.
The problems that result can be quantified
in the case of the isovector vector channel, where the hadronic
spectral function is known experimentally.
As shown in Ref.~\cite{kmfesr}, 
the errors in integer-power weighted FESR's, even at
scales as high as $m_\tau^2$,
can be very large, despite the fact that the OPE at this scale is both
dominated by the leading ($D=0$) perturbative term, and 
rather rapidly converging.  (A more detailed discussion of why it is that
one cannot judge the degree of
validity of the local duality approximation simply by 
looking at the degree of convergence of the OPE in a given region
is given in Ref.~\cite{kmfesr}.)

These problems, encountered in employing
the conventional, integer-power-weighted version of
FESR's, are, however, not intrinsic to the FESR approach.  Indeed,
there is at least one example of a non-integer-power-weighted FESR that
is very well-satisfied: that giving the hadronic $\tau$
decay widths in terms of an integral over the circle of radius
$s_0=m_\tau^2$ of the product of the OPE for the isovector vector
current correlator and the weight function 
$w_\tau (s)=\left( 1-s/m_\tau^2\right)^2 \left( 1+2s/m_\tau^2\right)$
(where the dominant input parameter in the OPE representation is
$a(m_\tau^2)$, which can be taken as obtained by running the
value measured at the $Z$ mass down to the $\tau$ scale).
The reason for the success of this sum rule is well-known: the
juncture of the cut and circular portions of the FESR contour
corresponds to the edge of hadronic phase space and hence,
because of kinematics, the weight function $w_\tau (s)$ has a
(double) zero at $s=m_\tau^2$.  This zero suppresses contributions
from the portion of the circular part of the contour near the
real timelike axis for which the OPE representation of the correlator
is not reliable (at intermediate scales like $m_\tau^2$)\cite{taurefs}.
This immediately suggests that the appropriate way to implement
FESR's in other
channels is, not with integer-power weights, but rather with
weights having a zero at $s=s_0$.  In Ref.~\cite{kmfesr} it
was shown that, in the isovector vector channel, where one can
check the procedure explicitly, weight functions of either the form
\begin{equation}
w_s(s)=\left( 1-{\frac{s}{s_0}}\right) \left( 1+A{\frac{s}{s_0}}\right) \ ,
\label{singlepinch}
\end{equation}
having a single zero at $s=s_0$, or
\begin{equation}
w_d(s)=\left( 1-{\frac{s}{s_0}}\right)^2 \left( 1+A{\frac{s}{s_0}}\right) \ ,
\label{doublepinch}
\end{equation}
having a double zero at $s=s_0$, produce extremely well-satisfied 
FESR's for a wide range of values of $s_0$ and
the continuous parameter, $A$.  In addition,
using {\it only} the OPE representation, for a range of $A$ and $s_0$,
and fitting the parameters of a sum-of-resonances ans\"atz
to this representation, results in a very good reconstruction
of the hadronic spectral function, including a determination
of the $\rho$ decay constant accurate to within a few $\%$\cite{kmfesr}.
In order
to have a compact terminology for use in describing the weight
families in Eqs. (\ref{singlepinch}) and (\ref{doublepinch}), 
we will refer to $w_s(s)$ and $w_d(s)$ as single-pinch
and double-pinch weights, respectively.
The freedom to vary $A$ plays a role analogous to that of the
variation of $M$ within the stability window in an SVZ-style analysis,
in that it strengthens the sum rule constraints on the parameters
of the spectral ans\"atz by working with a range of different 
weight profiles.  An additional advantage, at least if one wishes to
determine not just the parameters of the lowest
resonance in the channel but also those associated with higher
resonances, is that the weight function
can actually be arranged to be larger in the second resonance
region than in the first.

In what follows, in light of its success in
the isovector vector channel, we will employ the FESR framework, using both
the single- and double-pinch weight families, to investigate
the isospin-breaking vector current correlator, 
$\Pi^{38}$, defined above.  As usual in the FESR framework,
we will work at scales as high as possible, compatible with
the constraint of having a tractable and physically sensible
spectral ans\"atz for $s<s_0$.  
Since little is known about the vector meson resonance spectrum
beyond the second excited resonance region, and since including even
the second excited resonance region would lead to a spectral
ans\"atz with more parameters than are generally tractable for the present
analysis, we are forced
to work at scales no higher than $\sim 2.8$ GeV$^2$.
Since the separation of the
first and second excited vector meson resonance regions 
is comparable to the resonance widths
(the $\rho^\prime$ and $\omega^\prime$ lie at $1419$ and $1452$
MeV, the $\rho^{\prime\prime}$ and 
$\omega^{\prime\prime}$ at
$1723$ MeV and $1649$ MeV, respectively\cite{pdg98}),
it is clear that, at these scales, we are not yet in the
region of the validity of local duality.  This makes the
use of the single- and double-pinch families crucial to
the reliability
of the FESR analysis.  
In order to maintain as good convergence
as possible on the OPE side of the two sum rule families, while at the
same time allowing enough variation in $s_0$ to get a
good determination of the parameters of the spectral ans\"atz,
we also restrict our attention to scales, $s_0$, greater
than $2$ GeV$^2$.  
%At the lower end of this $s_0$ range,
%$D=6$ contributions on the OPE side
%become non-negligible.
%Normally this would limit the reliability of the
%analysis since, in the absence of detailed phenomenological
%information on the full range of $D=6$ vacuum condensates,
%one is usually forced to ``estimate'' their
%values using the vacuum saturation approximation\cite{svz}.  This
%approximation allows all such condensates to be written in
%terms of the square of the quark condensate, but is known
%to be rather crude, in cases where the combined $D=6$
%contribution has been obtained phenomenologically\cite{???}.
%Fortunately, it turns out that the 
%ability to employ both of the weight families
%discussed above allows one to place self-consistency constraints
%on the relevant combination of $D=6$ condensates, and hence
%reduce the uncertainties on the extracted spectral parameters.

\section{Details of the Analysis}
Since the general framework to be employed in the analysis has been outlined
in the previous section, it remains only to discuss
the input required on the hadronic and OPE sides
of the various sum rules.

We begin with the hadronic side.
We take, as our ans\"atz for the hadronic spectral function, a sum
of resonance contributions.  For the scales used in the
analysis, the resonances present in the region of
the hadronic spectral integral are
the $\rho$, $\omega$, $\phi$, $\rho^\prime$ and $\omega^\prime$.
(Although the tails of the $\rho^{\prime\prime}$ and 
$\omega^{\prime\prime}$ intrude slightly into the hadronic
integration region for $s_0$ near $2.8$ GeV$^2$, 
their contributions are strongly suppressed
by the zeros in the weight functions.  We have checked that
including an effective, combined 
$\rho^{\prime\prime}$-$\omega^{\prime\prime}$ contribution in
the spectral ans\"atz has negligible effect on the extracted
$\rho$, $\omega$ and $\phi$ spectral strength parameters.) 
We thus include contributions, written in terms of
Breit-Wigner resonance forms,
for all these resonances.  Because the separation of the
$\rho^\prime$ and $\omega^\prime$ is much smaller than either
of their widths, and also to reduce the number of free parameters
in the spectral ans\"atz, we have combined the latter two contributions.
The strong overlap of the two resonances would,
in any case, prevent one from being able to sensibly extract separate
$\rho^\prime$ and $\omega^\prime$ strength parameters by means of
any sum rule analysis of $\Pi^{38}$.

The spectral ans\"atz
then has the form
\begin{equation}
\rho^{38} (q^2)=
{\frac{1}{4\sqrt{3}}}\left[ f_\rho \hat{\delta} \left( q^2-m_\rho^2\right)
-f_\omega \hat{\delta} \left( q^2 -m_\omega^2\right) + 
f_\phi\hat{\delta} \left( q^2 -m_\phi^2\right)
+f_{\rho^\prime\omega^\prime}\delta \left( q^2-
\bar{m}_{\rho^\prime\omega^\prime}^2\right)\right]\ ,
\label{specansatz}
\end{equation}
where 
\begin{equation}
\hat{\delta}\left( q^2-m^2\right)\equiv {\frac{1}{\pi}}
{\frac{m\Gamma}{\left( q^2-m^2\right)^2+m^2\Gamma^2}}
\label{wideresonanceapprox}
\end{equation}
(with $\Gamma$ the width of the resonance in question).
This expression reduces
to $\delta \left( q^2-m^2\right)$ in the narrow width approximation (NWA).
The minus sign in front of $f_\omega$ and the factor of 
$1/4\sqrt{3}$ are
conventional; inclusion of the former ensures
that $f_\omega$ and $f_\rho$ become equal
in the limit that the spectral contributions in the $\rho$-$\omega$ region
are generated entirely by leading order $\rho$-$\omega$ mixing.
For the combined $\rho^\prime$-$\omega^\prime$ contribution we
have taken average values for the effective mass and width.
$f_\rho$, $f_\omega$, $f_\phi$ and $f_{\rho^\prime\omega^\prime}$
are free parameters, to be determined from the matching of
hadronic and OPE sides of the single- and double-pinch sum rules,
for a range of $s_0$, $A$ values.

A few comments are in order concerning the form
of the ans\"atz above, and the physical meaning of the
parameters to be extracted from the analysis which follows.

The first concerns the 
need for the inclusion of a $\phi$ contribution.
Note that the correlator $\Pi^{38}$ is very closely related to that,
$\Pi^{\rho\omega}$, obtained by dropping the strange part of
the hypercharge current from $\Pi^{38}$ (the OPE's for
the two correlators are, in fact, identical to three-loop order). 
The latter correlator has been
studied in a number of earlier SVZ-style analyses\cite{svzro,hhmk,kmro,ijl}.
In the earliest of these, the NWA
was employed for all resonances, and no $\phi$ contribution was included in
the spectral ans\"atz\cite{svzro,hhmk}.  As pointed out
in Ref.~\cite{kmro}, however, the existence of significant cancellations
between the NWA $\rho$ and $\omega$ contributions (which would be
exact in the limit of mixing dominance and equality of $\rho$
and $\omega$ masses) means that a $\phi$ contribution,
even if significantly smaller than the {\it individual} $\rho$
and $\omega$ contributions, could nonetheless be important.
Performing the sum rule analysis with a $\phi$ contribution included
shows that this is, indeed the case\cite{kmro}.  
Including the $\phi$ contribution
in the spectral ans\"atz also cured an unphysical feature of
the solutions obtained earlier, which did not include it\cite{kmro}.
The analysis of Ref.~\cite{kmro}, 
however, still employed the NWA
for all resonances.  

The second point concerns the
need to incorporate the $\rho$ width into the analysis.
Because, again, of the high degree of
cancellation between the NWA $\rho$ and $\omega$ contributions,
it was pointed out that  
the precise degree of this cancellation might well be
sensitive to whether or not the difference between
the $\rho$ and $\omega$ widths 
was retained in the spectral ans\"atz\cite{ijl}.  
The subsequent analysis of Ref.~\cite{ijl} showed that this is, indeed,
the case:  the spectral parameters,
$f_V$, decrease by factors $\sim 6$ when one employs the
physical widths in place of the NWA.

The third point concerns the interpretation of the higher resonance
strength parameters, $f_\phi$ and $f_{\rho^\prime\omega^\prime}$.
It is, of course, very natural to take the spectral function
to be resonance dominated.
Moreover, the near-threshold region of the spectral function has been
computed to two loops in ChPT\cite{km2loop},
and one can see from this result that the corresponding low-$s$ background
contribution to the relevant spectral integrals is
tiny compared to that from the $\rho$-$\omega$ region.  The
case of the $\phi$, however, is less clear, since background
contributions above the $\rho$-$\omega$ region are
not amenable to as reliable estimates, and hence might not
be similarly negligible.  In the ans\"atz as written,
such physical contributions, if present, could only be mocked up
(approximately) by additional effective contributions to the 
$\phi$ and $\rho^\prime$-$\omega^\prime$ strengths.  Thus, one
must use some caution in interpreting, for example, the
extracted $f_\phi$
in terms of the physical resonance parameters $F^3_\phi$ and $F^8_\phi$ -- 
some portion of $f_\phi$ could actually correspond
to an averaged version of background contributions in the region
between $\rho$-$\omega$ and $\rho^\prime$-$\omega^\prime$.  
The quality of the agreement between the hadronic and OPE sides
of our sum rules is, however, {\it post facto} evidence in
favor of resonance dominance, and hence also in
favor of the possibility of interpreting
$f_\phi$ in terms of $F^3_\phi$ and $F^8_\phi$.

Let us turn, then, to the input on the OPE side of the sum rules.  We
will discuss the contributions, in turn, by operator dimension.

Since the correlator in question is isospin-breaking, the only dimension
$D=0$ contribution to $\Pi^{38}$ is electromagnetic
(we adhere, here, to common usage,
according to which the leading mass-dependent perturbative terms
are labelled $D=2$).
We retain
only the leading order (2-loop) graph in this case.  

The $D=2$ contributions are dominated by the strong interaction
terms proportional to $(m_d-m_u)^2$.
To 3-loop order, the results for these terms follow from the 3-loop
expressions for the correlator involving a flavor-non-diagonal
current and its conjugate\cite{ck93}, since
the perturbative contributions involving
two quark loops and a purely gluonic intermediate state
(present for flavor diagonal
currents but not for flavor-non-diagonal currents)
do not enter until 4-loop order.  The resulting expressions
are given in the Appendix.  To evaluate them, we require the
running masses, $m(Q^2)$, and running strong coupling,
$\alpha_s(Q^2)$.  These can be obtained once the values are determined
at any fixed scale, $\mu_0$.  Since the 4-loop $\gamma$\cite{gamma4} and 
$\beta$\cite{beta4}
functions for QCD are now known, we have employed these
when running the masses and coupling (explicitly, we 
solve the RG equations exactly, using the truncated 4-loop
$\gamma$ and $\beta$ functions as input).

As input for the running coupling, we take $\mu_0 =m_\tau$ and
use the latest (1998) value for $\alpha_s(m_\tau^2)$
obtained by the ALEPH Collaboration
in their analysis of non-strange hadronic $\tau$ decays\cite{ALEPH98}.
(The analysis of the strange
decays employed previous theoretical results for the $D=2$ terms,
proportional to $(m_s-m_u)^2$, which turn out to be in 
error\cite{kmmsprob,chetyrkin,pp}; the value obtained in the global
ALEPH analysis must, therefore, be excluded.)

The situation with the light quark mass
difference $\delta m\equiv (m_d-m_u)$ is somewhat more complicated.
We first write
\begin{equation}
\delta m=\left({\frac{m_d-m_u}{m_d+m_u}}\right) (m_d+m_u)
\equiv r(m_d+m_u)\ .
\label{delm}
\end{equation}
The isospin-breaking mass ratio, $r$, is known, from a number 
of ChPT analyses, to be $r=0.288\pm 0.037$\cite{leutwylermq}, which
would allow one to determine $\delta m$ if $m_d+m_u$ were
known.
The most recent determination of $m_d+m_u$
is that based on an integer-power-weighted FESR analysis of the 
isovector pseudoscalar channel\cite{bpr,prades97}.
In this analysis, the pion pole contribution to the spectral function
is known experimentally but the
continuum contribution is not.  The authors of Refs.~\cite{bpr,prades97},
therefore, constructed an ans\"atz for the unmeasured
continuum contribution.  It turns out
that the continuum portion of the resulting model
spectral function provides
roughly $3/4$ of the contribution to the extracted
value of $(m_d+m_u)^2$.
Unfortunately, it has recently been shown, using
the FESR framework discussed above, that this continuum ans\"atz
is unphysical\cite{kmfesr}, so one cannot employ the values of
Refs.~\cite{bpr,prades97}.  

If $m_s$ were known (at some scale),
then one could straightforwardly determine $m_d+m_u$ (at that same 
scale) using the known
(scale-independent) ratio of masses, $r_s=2m_s/(m_d+m_u)=24.4\pm 1.5$, 
obtained by Leutwyler\cite{leutwylermq} using ChPT.
Unfortunately, the situation is also somewhat
complicated for $m_s$.  A number of recent analyses produce values
of $m_s (1\ {\rm GeV}^2)$ (in the $\overline{MS}$ scheme)
ranging from $\sim 110$ MeV to $\sim 210$ MeV, often with rather
large errors\cite{jm,cps,cfnp,narison3388,km3388,kmmsprob,chetyrkin,jnew}.
Because the analyses based either on flavor breaking in hadronic $\tau$ 
decays\cite{kmmsprob,chetyrkin} or Narison's
$\tau$-decay-like sum rule for 
$\Pi^{33}-\Pi^{88}$\cite{narison3388,km3388} 
have rather large errors resulting
from experimental uncertainties which are unlikely to be significantly
improved in the near future,
the most favorable approach would appear to be that
based on various sum rule treatments of the strange scalar
channel, where the dominant $K\pi$ part of the spectral function
is in principle determined, via the Omnes representation of the timelike
scalar $K\pi$ form factor, in terms of experimental $K\pi$ phase
shifts and $K_{e3}$ data\cite{jm}.  The most recent analyses of
this channel\cite{cfnp,jnew} employ the SVZ framework, and
produce values 
$m_s(1\ {\rm GeV}^2)=125-160$ MeV\cite{cfnp}, and
$160\pm 30$ MeV\cite{jnew}.  (The same low-$s$ part
of the spectral function is
used in both analyses; the only difference between the two lies
in the treatment
of the ``continuum''.  The results of Ref.\cite{jnew}, 
in addition, show no stability window
for $m_s$).  Preliminary
work using the FESR framework discussed above indicates
that residual errors associated with the use of the local duality 
approximation in the continuum region
remain, for this channel, when one uses the SVZ approach.  
(See, e.g., the results of Ref.~\cite{kmfesr}.  From these one can see
(1) that using the central values for the parameters describing
the fit to the $K\pi$ phases from Refs.~\cite{jm,cfnp}, together with
the central value from the $m_s$ range from Ref.~\cite{cfnp},
one obtains rather poorly satisfied families of FESR's, and (2) that using
the spectral function of Refs.~\cite{jm,cfnp,jnew},
again with central values for the fit parameters, the FESR analysis, in fact,
produces $m_s$ values larger by $\sim 20$ MeV than those obtained
in the analysis
of Ref.~\cite{cfnp}).  Although work on the extraction
of $m_s$ is still in progress\cite{mgb},
we conclude already from the preliminary results noted above, that
$m_s(1\ {\rm GeV}^2)\sim 165$ MeV, probably with errors 
$\sim \pm 20$ MeV or less.  The ChPT ratio then produces
$(m_d+m_u)(1\ {\rm GeV}^2)\simeq 13.5$ MeV, with errors $\sim\pm 2$ MeV.
For any value in this range it turns out that the $D=2$ contributions
are at the $\sim 15\%$ (or less) level of the $D=4$ contributions, and
the resulting errors lead, therefore, to very small ($\%$ level)
uncertainties in the final results.  Since these uncertainties
are much smaller
than those generated by the uncertainty in the isospin-breaking mass
ratio $r$, we have employed the central value 
$(m_d+m_u)(1\ {\rm GeV}^2)=13.5$ MeV, and retained only the uncertainty
in $r$, in the analysis which follows.

The $D=4$ contributions are much more straightforward.  
Although in principle both those $D=4$ terms
proportional to the isospin-breaking mass difference,
$\delta m$, 
and those proportional to the isospin-breaking condensate difference,
$<\bar{d}d>-<\bar{u}u>$, appear in the OPE of $\Pi^{38}$,
the latter are numerically tiny compared to the former.
The dominant $D=4$ contribution can then be written in terms 
of $r$ and the combination $(m_d+m_u)<\bar{q}q>$, which we
can take from
the GMO relation
\begin{equation}
(m_d+m_u)<\bar{q}q>=-m_\pi^2f_\pi^2\ .
\label{gmo}
\end{equation}
The dominant uncertainties for the $D=4$ terms thus result from
those on $r$.

The phenomenological situation is not so favorable in the
case of the $D=6$ condensates.  Usually, in the absence
of pre-existing determinations of the relevant condensates, one
makes estimates based on the vacuum saturation approximation (VSA).
It is well known that, in situations where it has been possible to
perform phenomenological checks by extracting the total
$D=6$ contribution from data, the VSA has proven to 
significantly underestimate
these contributions\cite{vacsat}.  Usually one simply replaces
the factor $\alpha_s<\bar{q}q>^2$, which is produced by the VSA,
by an effective scale-independent factor, written
$\rho^\prime\alpha_s<\bar{q}q>^2$.  The parameter, $\rho^\prime$, then 
represents the deviation from the VSA.  Ideally, it should either
be possible to determine $\rho^\prime$ from data, or the $D=6$
contributions should be small, for the sum rule in question.
In our case, neither of these conditions holds.  In particular,
because we are forced to work at scales as low as $2$ GeV$^2$
in order to constrain the spectral parameters, the $D=6$
contributions can, for $s_0\sim 2$ GeV$^2$, and
certain values of $A$ employed in our analysis, approach $\sim 40\%$
of the leading $D=4$ term.  Fortunately, it turns out, as
we will see explicitly below, that by working with both the
single- and double-pinch weight families, we can actually
obtain a rather good determination of the $D=6$ contribution
to the correlator (albeit it as
a function of $r$) by insisting on the consistency
of the results obtained from the two different sum rule 
families.  

In the Appendix, it is shown that the VSA leads
to an expression for the $D=6$ contribution to $\Pi^{38}$
proportional to
\begin{equation}
\alpha_s \left(\langle  \bar{d}d\rangle^2
-\langle  \bar{u}u\rangle^2\right)\ =\ 
\gamma \left( \alpha_s \langle \bar{q}q\rangle^2\right)\ ,
\label{vacsat}
\end{equation}
where $\langle \bar{q}q\rangle$ is the average of the $u$ and $d$
condensates, and 
\begin{equation}
\gamma\equiv {\frac{\langle \bar{d}d\rangle}{\langle \bar{u}u\rangle}}-1\ ,
\label{gamma}
\end{equation}
describes isospin-breaking in the light quark condensates.  In order
to compare the deviation from the VSA in the isospin-breaking channel
with that in the analogous isospin-conserving isovector vector ($ab=33$)
channel, we write the re-scaled version of the RHS of Eq.(\ref{vacsat})
in the form
\begin{equation}
\rho^\prime\alpha_s \left(\langle  \bar{d}d\rangle^2
-\langle  \bar{u}u\rangle^2\right)\ =\
\rho_{red}\gamma \left(\rho \alpha_s \langle \bar{q}q\rangle^2\right)\ ,
\label{defnrhored}
\end{equation}
where $\rho$ is the parameter describing the deviation from the VSA in
the $33$ channel, and one has, phenomenologically,\cite{narisonrho}
\begin{equation}
\rho \alpha_s \langle \bar{q}q\rangle^2 = (5.8\pm 0.9)\times 10^{-4}
\ {\rm GeV}^6\ .
\label{value33rho}
\end{equation}
With this definition, $\rho_{red}$ reduces to $1$ 
in the limit that the deviation from the VSA is the same in the $33$ and
$38$ channels.

The consistency procedure for fixing the 
$D=6$ contribution to $\Pi^{38}$, together with the phenomenological
input of Eq.~(\ref{value33rho}), of course, 
determines only the
product $\rho_{red}\gamma$.
In 
presenting our results for $\rho_{red}$ below, we have taken
$\gamma \simeq -0.008$, which represents an average of
the previous determinations listed in Ref.\cite{hhmk}, bar one.
(We omit the value based on an analysis of
baryon splittings because it implies
(via the 1-loop ChPT relations between flavor-breaking and
isospin-breaking in the light quark condensate\cite{gl85}) 
$\langle \bar{s}s\rangle /  \langle \bar{u}u\rangle >1$, which
appears unphysical).
We will discuss the determination of $\rho_{red}$ in more detail below
when we present the results of the analysis.

The last point in need of discussion concerns the way in which we handle
the integrals 
on the OPE side of the various FESR's.
Two options exist in the literature.  The first, sometimes
called the ``fixed order expansion'', involves first
expanding $\alpha_s(Q^2)$ and the
mass factors, generically $m(Q^2)$, in terms of
$\alpha_s(s_0)$ and $m(s_0)$.  The coefficients of the
perturbative expansions in powers of $\alpha_s(s_0)$ 
are then polynomials in $\ell n(s/s_0)$\cite{kniehl},
and the desired contour integrals can thus be
written in terms of
elementary integrals involving logarithms and powers of $s$.
The integrated OPE expressions which result
involve $m(s_0)^2$, multiplied by a power series in $\alpha_s(s_0)$.
There is, of course, in this expression, the usual residual
dependence on the choice of scale $s_0$ for the expansions
discussed above, which results from truncating the full 
perturbative series at fixed order.  The second alternative,
often referred to as ``contour improvement'',
involves numerically integrating the factors
$\left[ m(Q^2)\right]^k \left[\alpha_s(Q^2)\right]^j s^\ell$
around the circular contour in the $s=-Q^2$ plane\cite{pledib}.
It is known that this has the effect of simultaneously
improving the convergence of the perturbative series and
reducing the residual scale dependence\cite{pledib,bpr,prades97}.
As a result, we have evaluated all the integrals on the OPE
sides of our sum rules using this approach.

Let us now turn to the results, which are presented in the Table.
As explained above, the dominant uncertainty is due to that
in the ChPT determination of $r$.  We have, therefore,
tabulated the results for the range of values corresponding
to the errors on $r$ quoted by Leutwyler\cite{leutwylermq}.
All results are based on matching the hadronic and OPE sides 
of the two sum rule families for $s_0$ in the range $2.0$ to $2.8$
GeV$^2$, and with $A$ in the range $2$ to $5$ for the 
single-pinch case and $3$ to $6$ in the double-pinch case.
The choice of range of $A$ in each case
has been made so as to keep the convergence of
the perturbative series for the $D=2$ term under control.
It is worth mentioning that the quality of the match between the OPE
and hadronic sides which results after the fitting of the spectral parameters
is significantly better for Leutwyler's central value of $r$.

The value of the $D=6$ VSA-violating parameter, $\rho_{red}$,
given in the Table, is determined by
requiring that the valueds of $f_\rho$ obtained using
the single- and double-pinch weight families are the same.
The sensitivity of
$f_\rho$ to variations in $\rho_{red}$
(true also of the other $f_V$), as well
as the difference in the $\rho_{red}$ dependence of $f_\rho$
for the single- and double-pinch analyses,
is shown in Figure 2.
The fact that,
once $\rho_{red}$ has been determined by the requirement
of the consistency of the two output $f_\rho$ values, all the rest of the
spectral parameters, determined using either the single-
or double-pinch weights, also become consistent
is strong evidence in favor of the reliability of the analysis.
Note that (1) the possibility of determining the correction
to the VSA for the $D=6$ operators, and (2) the inclusion of both
the $D=2$ terms and the ${\cal O}(\alpha_s ,\alpha_s^2)$
contributions to the Wilson
coefficient of the $D=4$ term, are features not present in
previous analyses of the analogous isospin-breaking $\Pi^{\rho\omega}$
correlator.
Although the value of $\rho_{red}$, determined as just described,
depends somewhat on $r$, this dependence is not
strong, and we obtain $\rho_{red} =1.15\pm 0.15\pm 0.2$.  The first
error corresponds to that in Eq.~(\ref{value33rho}),
the second to that on $r$.  We see that the violation 
of the VSA is very similar in both the $33$ and $38$ channels.
The importance, in
reducing the errors on the determinations of the
spectral parameters, $f_V$, of being able to determine $\rho_{red}$
is also evident from Fig.~2.

Having determined the $D=6$ contributions by self-consistency,
the errors on the extracted values of $f_V$ are determined
solely by those on $r$, and are $\sim 10-15\%$, completely
correlated with $r$.  

Having extracted the parameters $f_V$, it is straightforward
to determine the isospin violating decay constants.  One finds
\begin{eqnarray}
F^8_\rho &=&2.4\pm 0.3\ {\rm MeV} \nonumber \\
F^3_\omega &=& -3.4\pm 0.4\ {\rm MeV}\nonumber \\
F^3_\phi &=& 0.33\pm 0.02\ {\rm MeV}\ ,
\label{decayconst}
\end{eqnarray}
where the errors reflect those on the input isospin-breaking mass 
ratio, $r$.

\section{Summary and Discussion of Phenomenological Consequences}
A number of useful general
observations follow from the analysis above.  First, we
have found that the violation of the
VSA for the $D=6$ condensates is very similar
in the isospin-breaking ($38$) and isospin-conserving
($33$) vector current channels. Second, we have seen that,
although the analysis
can be 
performed successfully for any $r$ in the range given
by Leutwyler, the central value of that range 
is preferred, in the sense of
giving the best match between OPE and hadronic sides of both the
single- and double-pinch families of
sum rules.  Finally, we have demonstrated that
the FESR
method, particularly when implemented using both the single- and
double-pinch weight families, is very effective, allowing
a determination of the $\rho^{38}$ spectral parameters, $f_V$,
with rather small errors.
These errors (between $10$ and $15\%$ for $f_\rho$,
$f_\omega$ and $f_\phi$) are a factor
of $3$ smaller than those obtained in the earlier analysis
of Ref.~\cite{krmspec} based on results of an SVZ
analysis of $\Pi^{\rho\omega}$\cite{ijl}.
In order to obtain this level of reduction,
the ability to self-consistently determine $\rho_{red}$
was crucial.

Let us now turn to the phenomenological consequences of our
results.  First note that the corrections required
to convert the measured contribution of the vector meson, $V$, 
to the EM spectral function, $\rho^{EM}$,
into the corresponding contribution to either $\rho^{33}$ (for $V=\rho$)
or $\rho^{88}$ (for $V=\omega ,\ \phi$) are given by the ratios
\begin{eqnarray}
\left[ \frac{F^{3}_\rho}{F^{EM}_\rho}\right]^2&=& 0.982\pm 0.0021
\nonumber \\
\left[ \frac{F^{8}_\omega}{\sqrt{3}\, F^{EM}_\omega}\right]^2&=& 
1.154\pm 0.017 \nonumber \\
\left[ \frac{F^{8}_\phi}{\sqrt{3}\, F^{EM}_\phi}\right]^2&=& 
1.009\pm 0.001\ ,
\label{corrections}
\end{eqnarray}
where the numerical values follow from those in Eq.~(\ref{decayconst}).
The size of the deviations of the $\rho$ and $\omega$ 
corrections from $1$ are reduced by 
$\sim 15-20\%$ from those obtained in the earlier 
analysis \cite{krmspec}; that for the $\phi$ is increased, but remains
small.  In all cases the errors have been reduced by
a factor of $3$ or more.
Note that the first of these corrections is the one relevant to precision
tests of CVC.  Note also that, as claimed above, the corrections 
given in Eqs.~(\ref{corrections}),
for both the $\rho$ and $\omega$, lie in
the corresponding ranges produced by the estimate of Section II.

With the results given in Eq.~(\ref{corrections}), it is now
possible to correct the EM data used as input to the
inverse moment chiral sum rule for the $6^{th}$ order
LEC, $Q$.  The sum rule is given by\cite{gk95,gkinv}
\begin{eqnarray}
&&\int_{4m_\pi^2}^\infty  ds \ \frac{(\rho^{33}-\rho^{88})(s)}{s}
=
{\frac{16 (m_K^2-m_\pi^2)}{3F^2}} Q (\mu^2)
+ {\frac{1}{48 \pi^2}} \log \left( {\frac{m_K^2}{m_\pi^2}}\right) \nonumber \\
&&\qquad\qquad 
+\left(\frac{L_9^r(\mu^2)+L_{10}^r(\mu^2)}
{2 \pi^2 F^2}\right) \left[ m_\pi^2 \log \left({\frac{m_\pi^2}{\mu^2}}\right) 
- m_K^2\log\left({\frac{m_K^2}{\mu^2}}\right)\right]\ ,
\label{invchmom}
\end{eqnarray}
where $\mu$ is the renormalization scale of the effective chiral
theory and $L_k^r$ are the usual renormalized $4^{th}$ order LEC's
of Gasser and Leutwyler\cite{gl85}.   
This sum rule was evaluated in Ref.~\cite{gkinv} using as input 
EM data for the isoscalar spectral function and
both EM and $\tau$ decay data
for the isovector spectral function.
The corrections above, required for the EM data, were not
considered in this analysis.  It is not clear, from our reading of
the discussion of Ref.~\cite{gkinv}, exactly what 
the relative weightings of
$\tau$ decay and EM data in the determination of
the $\rho^0$ contribution to the LHS above actually were.
Since, however, $\tau$ decay data is considerably
more precise than electroproduction data, we have assumed in what
follows that the determination is dominated by
$\tau$ decay data.
To the extent that this is true,
we need only make corrections to the (nominally) isoscalar
$\omega$ and $\phi$ contributions.  The result of this exercise
is a shift of $Q(m_\rho^2)$ from $(3.7\pm 2.0)\times 10^{-5}$
to 
\begin{equation}
Q(m_\rho^2)=(2.4\pm 2.0)\times 10^{-5}\ .  
\label{Q3388value}
\end{equation}
(For reference, making, instead, the somewhat perverse
assumption that the determination of the $\rho^0$ 
contribution was dominated by
EM data, one would find $Q(m_\rho^2)=(2.0\pm 2.0)\times 10^{-5}$.
The full correction is dominated by that to the $\omega$ contribution.
The reason this correction is so much larger than the others
has been discussed above.)

One should bear in mind, in
interpreting these results, that (1) there are, in principle,
additional corrections to be made to the nominal isovector
and isoscalar contributions at higher $s$, and (2) the $\bar{K}K2\pi$
contributions were taken to be purely isoscalar in the analysis
of Ref.~\cite{gkinv}.  Because the separations within the higher
isovector and isoscalar resonance pairs,
$\rho^\prime$-$\omega^\prime$ and 
$\rho^{\prime\prime}$-$\omega^{\prime\prime}$, are much smaller than
the resonance widths, it is not possible to use sum rule methods to
extract the individual isospin-breaking decay constants of
these resonances.  As such, we are unable to estimate the size
of the former corrections.  Were the $\bar{K}K2\pi$ states to
have a significant isovector component, the effect would be
to raise $Q(m_\rho^2)$. 

An alternate method for determining $Q$ is
based on the observation that $Q$
occurs not only in the inverse chiral moment sum rule above, 
but also in the 2-loop ChPT expression for
$\Pi^{38}(0)$\cite{km2loop}.  
It is, thus, possible to make an
independent estimate by using the fitted spectral
ans\"atz for $\Pi^{38}$ to compute $\Pi^{38}(0)$, assuming
negligible contribution from the portion of the spectrum
above $2.8$ GeV$^2$.  One obtains, from this exercise,
\begin{equation}
Q(m_\rho^2)=(3.3\pm 0.4)\times 10^{-5}\ .  
\label{Q38value}
\end{equation}
Since there is no positivity constraint on $\rho^{38}(s)$
one does not know in which direction this result would be changed by
corrections due to the small higher-$s$ part of the spectral
integral.  The results of a study
of the effect of including two combined spectral contributions,
one for the $\rho^{\prime}$-$\omega^{\prime}$ and one for the
$\rho^{\prime\prime}$-$\omega^{\prime\prime}$, however, shows
negligible change in $Q(m_\rho^2)$, 
suggesting that
such corrections are unlikely to be numerically significant.
Since the two independent determinations of
$Q$ are completely consistent, within errors, 
the conclusions that $Q(m_\rho^2)\simeq 3\times 10^{-5}$
is considerably strengthened.

The last phenomenological application of our results concerns
the effect on Narison's $\tau$-decay-like sum rule for $m_s$.
Since a detailed discussion of the way in which one implements
the isospin-breaking corrections is given in Ref.~\cite{km3388},
we report here only the results of employing the improved
determinations of the correction factors determined above.
In doing so we will also take the opportunity
to update the input parameters to
the analysis of Ref.~\cite{km3388}, employing the newer (1998)
ALEPH value of $\alpha_s(m_\tau^2)$\cite{ALEPH98}.  
One finds, for example, using
$\tau$ decay data for the isovector input, that the average over
the values of $m_s(1\ {\rm GeV}^2)$ extracted using the set of
scales $s_0=1.4$, $1.5$ and $1.6$ GeV$^2$ in the sum rule analysis
is shifted from $138$ MeV to
$146$ MeV ($147$ MeV if one retains the 1997 ALEPH
value of $\alpha_s$ as input).  
Unfortunately, the errors on this value associated
with uncertainties in the experimental input are still very large,
$\sim \pm 50$ MeV at least, and this uncertainty cannot be appreciably
reduced without a significant improvement in the accuracy
of the determination of the experimental 
$\omega\rightarrow e^+e^-$ and $\phi\rightarrow e^+e^-$ widths.
As such, although the central value is brought into better
agreement with that discussed above, little more can learned
from the Narison sum rule, at the present time.

\acknowledgements
KM would like to acknowledge the ongoing support of the Natural
Sciences and Engineering Research Council of Canada, and to thank
Terry Goldman, Andreas H\"ocker,
Heath O'Connell, Derek Leinweber, Tony Thomas, and Tony Williams
for useful discussions.

\appendix
\section*{The OPE for $\Pi^{38}$}
The explicit form of the OPE for $\Pi^{38}$, keeping terms only up
to dimension six, and to ${\cal O}(\alpha_s^2, m_q^2)$, can
be obtained, as explained in the text, from the relevant expressions
for the flavor-non-diagonal case given in the 
literature (see Ref.~\cite{ck93} and the paper by Braaten, Narison
and Pich (BNP) in Ref.~\cite{taurefs}).  We list the results by
operator dimension.
\vskip .15in
\noindent
Dimension 0:\hfill
\vskip .075in
The only isospin-breaking contribution at dimension 0 is that due to EM,
and is given by\cite{svzro}
\begin{equation}
\left[ \Pi^{38}_{1\gamma E}\right]_{D=0}
= -{\frac{\alpha} {16\pi^3}}{\frac{1}{4\sqrt{3}}}ln(Q^2)
\end{equation}
where $\alpha$ is the usual EM coupling.
\vskip .15in
\noindent
Dimension 2:\hfill\vskip .075in
The $D=2$ term consists of the leading mass-dependent part
of the perturbative contribution to the OPE, and follows from the
expression given in Ref.~\cite{ck93}.  One finds
\begin{eqnarray}
Q^2\left[ \Pi^{38}(Q^2)\right ]_{D=2}&=&
{\frac{3}{2\pi^2}}{\frac{1}{4\sqrt{3}}}
\left[(m_d^2-m_u^2)(Q^2)\right]
\left [1+{\frac{8}{3}} a(Q^2)\right. \nonumber \\
&&\qquad \left. +\left (
{17981\over 432}+{62\over 27} \zeta(3)
-{1045\over 54} \zeta(5)\right ) a^2(Q^2)\right ]
\label{D2}
\end{eqnarray}
where $a(Q^2) = \alpha_s(Q^2)/\pi$, and $\zeta (n)$ is the Riemann zeta 
function.  Further details on how
the running of the coupling and 
the masses is handled can be found in Section IV.
\vskip .15in
\noindent
Dimension 4:\hfill\vskip .075in
Our expression for the $D=4$ contribution also follows from that
given in Ref.~\cite{ck93}.
Only the $m_q^4$ and quark condensate terms survive once one takes the
relevant isospin-breaking difference.  The former are numerically
tiny compared to the latter, and hence have not been written down
explicitly.  We then find
\begin{equation}
Q^4\left[\Pi^{38}(Q^2)\right]_{D=4} =
{\frac{2
\left(\langle m_u\bar{u}u\rangle - \langle m_d\bar{d}d\rangle\right)}
{4\sqrt{3}}}
\left[ 1+{1\over 3}a(Q^2)+{11\over 2}a^2(Q^2)\right] \ .
\end{equation}
The scale-invariant $<m\bar{q} q>$ difference 
can be written in terms of the $m_d-m_u$, $<\bar{d}d -\bar{u}u>$,
and the averages of the $u$ and $d$ quark masses and condensates.
Since isospin-breaking in the condensates is much smaller
than in the masses, the term proportional to $m_d-m_u$ dominates
numerically.  It can be recast in terms of the isospin-breaking
quark mass ratio, $r$, and $f_\pi$, $m_\pi$, as explained in
the text.
\vskip .15in
\noindent
Dimension 6:\hfill\vskip .075in
The 4-quark operators are the dominant operators at dimension 6.  Their
contribution to $\Pi^{38}$ can be obtained from the expressions given
in the Appendix of BNP\cite{taurefs}.
Since lack of phenomenological information on the various condensates
forces one to work with the re-scaled version of the VSA, one must,
for consistency, drop the terms of ${\cal O}(\alpha_s^2)$ contained
there.  (See the discussion of this point contained in BNP.)  One then finds
\begin{eqnarray}
4\sqrt{3}Q^6\left[ \Pi^{38}(Q^2)\right]_{D=6} &=&
-8\pi^2 a(Q^2)\left(\langle\bar{u}\gamma_{\mu}
\gamma_5 T^a u\bar{u}\gamma^{\mu}\gamma_5 T^a u\rangle -\langle
\bar{d}\gamma_{\mu}\gamma_5 T^a d\bar{d}\gamma^{\mu}\gamma_5 T^a d\rangle 
\right)\\
& & -{\frac{16\pi^2}{9}} a(Q^2)\sum_k
\left(\langle\bar{u}\gamma_{\mu}
T^a u\bar{q}_k\gamma^{\mu}T^a q_k\rangle -\langle
\bar{d}\gamma_{\mu}T^a d\bar{q}_k\gamma^{\mu}T^a 
q_k\rangle \right)\ ,
\end{eqnarray}
where $T^a$ is an $SU(3)$ generator.  Implementing the re-scaled VSA,
this expression reduces to
\begin{eqnarray}
4\sqrt{3}Q^6\left[ \Pi^{38}(Q^2)\right]_{D=6} &=&
{\frac{448\pi}{81}}\rho_{red}\gamma(\rho\alpha_s\langle\bar{q}q\rangle^2)\ ,
\end{eqnarray}
where $\gamma$, $\rho$ and $\rho_{red}$
are as defined in the text.

\vskip 1in\noindent
\begin{table}
\caption{Results for the $D=6$ VSA-violation parameter, $\rho_{red}$, and
the spectral strength parameters, $f_{\rho}$, $f_\omega$, $f_\phi$
and $f_{\rho^{\prime}\omega^{\prime}}$
as a function of the isospin-breaking mass ratio, $r$.  The first line, for
each value of $r$, corresponds to the results obtained using the 
single-pinch weight family, the second line to those obtained using
the double-pinch family.}
\vskip .3in\noindent
\begin{tabular}{cccccc}
$r$ & $\rho_{red}$ & $f_{\rho} (\times 10^3)$ & $f_{\omega} 
(\times 10^3)$ & $f_{\phi} (\times 10^3)$ & $f_{\rho^{\prime}
\omega^{\prime}} (\times 10^3)$\\ \hline
0.251 & 1.02 & 2.3 & 1.7 & -0.28 & -0.020 \\
& & 2.3 & 1.7 & -0.28 & -0.020 \\ \hline
0.288 & 1.15 & 2.6 & 2.0 & -0.32 & -0.026 \\
& & 2.6 & 2.0 & -0.32 & -0.026 \\ \hline
0.325 & 1.28 & 2.9 & 2.2 & -0.36 & -0.032 \\
& & 2.9 & 2.2 & -0.36 & -0.032 \\ 
\end{tabular}
\end{table}

\begin{figure} [htb]
\centering{\
\psfig{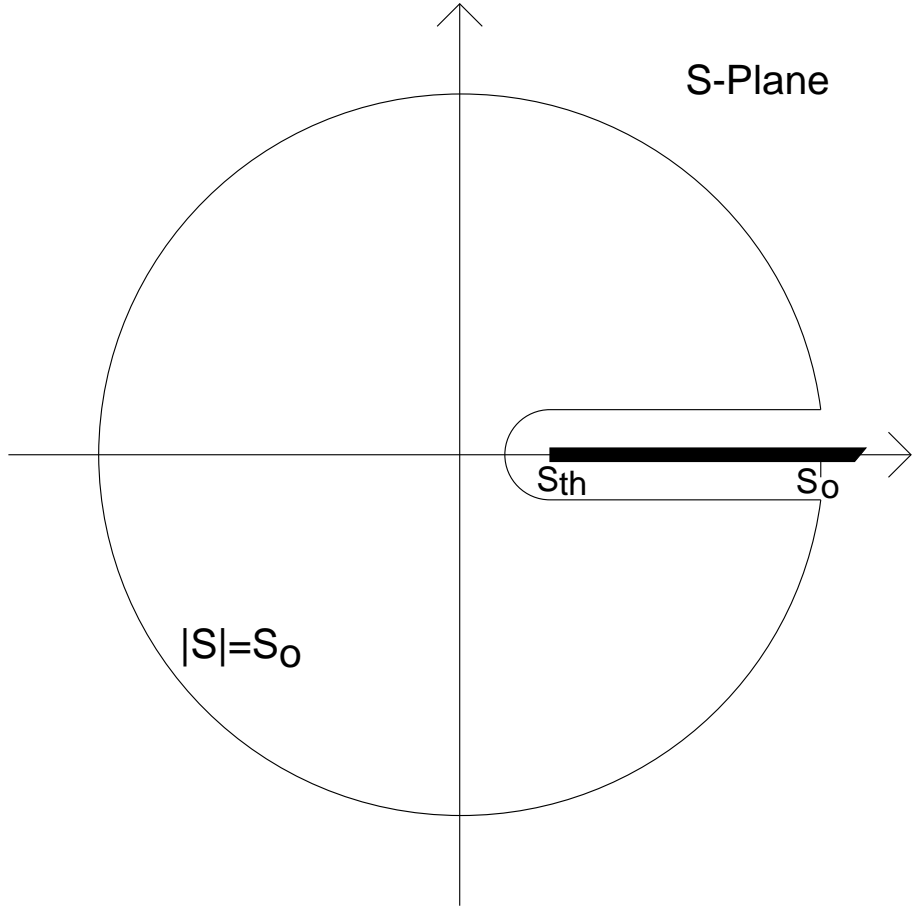}}
\vskip .5in         
\caption{The FESR ``Pac-man'' contour}
\label{figone} 
\end{figure}

\begin{figure}[htb]
  \centering{\
    \psfig{figure=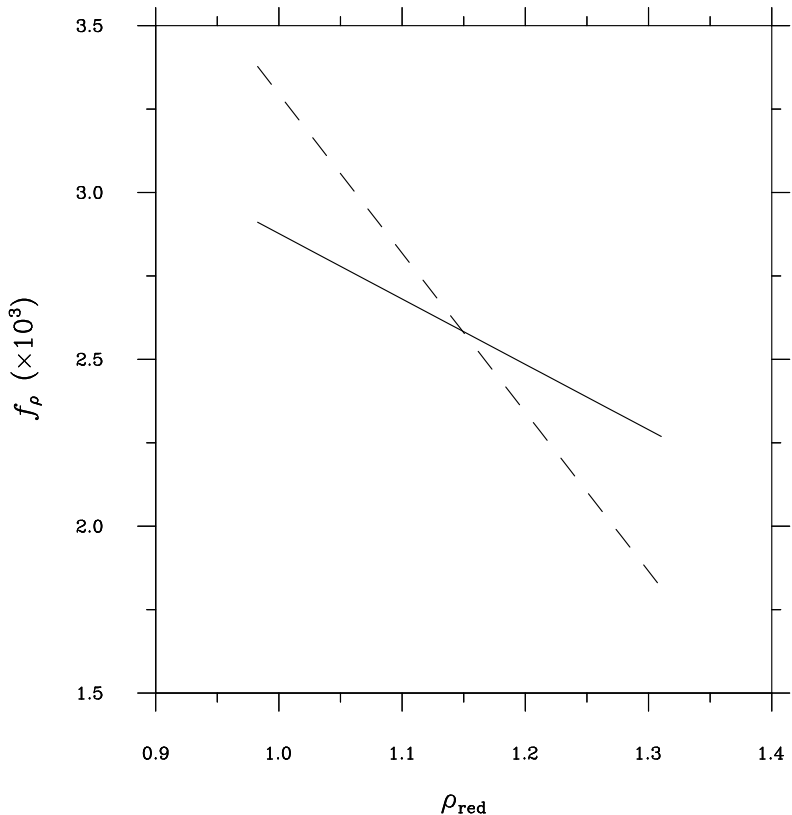,height=12.cm}}
\vskip .5in         
\parbox{130mm}{\caption
{The variation of $f_\rho$ with $\rho_{red}$ for the single-
and double-pinch weight families.  Results are displayed here
for the central value $r=0.288$.  The solid line corresponds
to the single-pinch weight analysis, the dashed line to the
double-pinch analysis.  The intersection point determines
the value of $\rho_{red}$ quoted in the table.}}
\label{figtwo}
\end{figure}

\end{document}